\documentclass[twoside]{article}

\usepackage{latexsym}      
\usepackage{graphicx}      
\usepackage{amsmath}
\usepackage{amssymb}
\usepackage[margin=1.0in,papersize={8.5in,11.0in}]{geometry}

\usepackage{fancyhdr}
\makeatletter
\renewcommand\section{\@startsection
   {section}{1}{0pt}%
   {-\baselineskip}%
   {0.1\baselineskip}%
   {\normalfont\large\bfseries}}%
\makeatother

\usepackage{caption}
\usepackage[compress,nospace]{cite}

\begin{document}

\begin{center}
{\Large\textbf{Blackhole evaporation model without information loss}}\\
\vspace{0.05in}
\textbf{Kristian Hauser A. Villegas}\\
\textit{National Institute of Physics\\ University of the Philippines Diliman, Quezon City 1101}\\
\textrm{khvillegas@gmail.com}\\

\vspace{0.15in}

\parbox{4.5in}{{\large \textbf{Abstract}}\\
\noindent{A simple model of a blackhole evaporation without information loss is given. In this model, the blackhole is \textit{not} in a specific mass eigenstate as it evaporates but rather, is in a superposition of various mass eigenstates and is entangled with the radiation. For astrophysical blackhole, the mass distribution is sharply peak about its average value with a vanishingly small standard deviation, which is consistent with our intuition of a classical object. It is then shown that as the blackhole evaporates, the evolution of the closed blackhole-radiation system is unitary. This is done by showing that the full density matrix satisfies Tr$\rho^2=1$ at all times. Finally, it is shown that the entanglement entropy, after an initial increase, decreases and approaches zero. These show that this model of blackhole evaporation has no infromation loss.}\\

\noindent{Keywords: blackhole, information paradox}}\\
\end{center}

\section{Introduction}
\label{sec:intro}
\hspace{\parindent}One phenomenon where the friction between the General Theory of Relativity and Quantum Mechanics is concretely felt is the information paradox in a blackhole evaporation\cite{Hawking76}. One can think of this process as follows. In the vicinity of the horizon, an entangled pair of particle and antiparticle is created, with one particle falling into the horizon while the other escaping into infinity as the Hawking radiation. As more pairs are created, the entanglement entropy associated with the reduced density matrix of the radiated particles, which is a mixed state, monotonically increases. The blackhole will eventually disappear as the evaporation is completed leaving a large entanglement entropy. In this picture, one therefore have a \textit{closed} system that started from a pure state, that is, the state of the matter before collapsing into the blackhole, and ends up with a mixed state with a large entanglement entropy. 

In Hawking's 1976 paper\cite{Hawking76}, he attributed this non-unitary evolution to the existence of a \lq\lq hidden surface\rq\rq, which is the horizon, in addition to the initial surface where data is given and the final surface where measurements are done. Various types of solutions have been proposed including string theory, non-locality, and complimentarity\cite{Susskind93a, Susskind93b, Lowe95} and a recent suggestion by Hawking et. al. that the information are stored in the horizon of the blackhole\cite{Hawking16}. I refer the reader to Hossenfelder's and Smolin's review article for a classification of these various solutions.

In this paper, I will take the conservative view that the non-unitary evolution is a consequence of combining the classical treatment of blackhole and the quantum-mechanical treatment of the radiation. Viewed this way, this non-unitary evolution is in fact related to the quantum measurement problem in quantum mechanics where the wavefunction of a quantum object undergoes a non-unitary collapse upon interaction with a classical measuring apparatus (see the review\cite{Schlosshauer05} for example). A simple quantum-mechanical description of the blackhole evaporation, that is, describing it in terms of a quantum state in a Hilbert space should therefore remove the information paradox. Specifically, the closed quantum system involving the blackhole, described by a state ket, and the radiation should evolve in a unitary way. In the next section, a model of blackhole evaporation is given. In this model, the blackhole is \textit{not} in a specific mass eigenstate as it evaporates but instead, is in a superposition of various mass eigenstates. In Section 2.1, a simple model is given for illustrative purposes. It is shown that during the blackhole evaporation, the entanglement entropy decreases back to zero after an initial increase. Lastly, the summary and conclusion is given.

\section{The Model}
In this discussion, I will denote the state of the nonrotating and uncharged blackhole by $|M\rangle$ where $M$ is the blackhole's mass. It is possible that there are other quantum numbers to completely specify the microstate of the blackhole. After all, it has entropy proportional to $M^2$. However, for my purpose, this is sufficient. In this notation, the state $|M=0\rangle$ means that the corresponding classical geometry is the Minkowski spacetime. 

The particle vacuum will be denoted by $|\Omega\rangle$. I will follow a simplified version similar to the the discussion by Mathur\cite{Mathur09} and denote the creation of neutral scalar particle-antiparticle pair at the horizon by the state
\begin{eqnarray}
\frac{1}{\sqrt{2}}|0\rangle |0\rangle +\frac{1}{\sqrt{2}}|1\rangle |1\rangle 
\end{eqnarray}
where $|0\rangle |0\rangle$ denotes a state without pair creation while $|1\rangle |1\rangle$ denotes a state with particle-antiparticle pair. The leftmost state in $|1\rangle |1\rangle$ is the ingoing particle that will cross the horizon, while the rightmost state is the outgoing particle that will escape to infinity. I will follow the specific choice of mode expansion by Hawking\cite{Hawking76} so that the outgoing particles can be interpreted as the particles detected by an observer with a worldline that is at a constant distance from the horizon.

In the original treatment of Hawking, the probability of creating $n$ particle-antiparticle pairs in the $j$th mode at the horizon is given by $P(n_j)=\frac{(1-x)(x\Gamma)^n}{[1-(1-\Gamma)x]^{n+1}}$ where $\Gamma=|t_{\omega}|^2$ is the transmission coefficient and $x=\exp\{-\omega T^{-1}\}$\cite{Hawking76}. In Eq.(1) on the otherhand, it is assumed for simplicity that there is only a single mode and that the relevant number of pairs created are only zero or one. Although Eq.(1) is a simplified version, it nevertheless captures the essence of pair creation at the horizon. Note also that if the particle-antiparticle pairs are fermionic, then Eq.(1) is just the coherent state $\sim\exp\{c^{\dagger}b^{\dagger}\}|\Omega\rangle$.

The process of pair creation at the blackhole horizon will then be denoted by
\begin{eqnarray}
|M\rangle |\Omega\rangle\rightarrow |M\rangle \bigg(\frac{1}{\sqrt{2}}|0\rangle |0\rangle +\frac{1}{\sqrt{2}}|1\rangle |1\rangle\bigg).
\end{eqnarray}

The pair creation process above is only possible in the presence of a blackhole, that is for $M>0$.

On the other hand, there is no pair creation on a flat spacetime so that
\begin{eqnarray}
|M=0\rangle |\Omega\rangle\rightarrow |M=0\rangle |\Omega\rangle.
\end{eqnarray}

I will now distribute the state $|M\rangle$ inside the parenthesis in Eq.(2) and consider its second term. One of the particle will fall into the blackhole while the other will escape into infinity. The falling particle will have a negative energy $-\delta m$ and will reduce the mass of the blackhole. We therefore have the process
\begin{eqnarray}
|M\rangle |1\rangle |1\rangle\rightarrow |M-\delta m\rangle |0\rangle |1\rangle
\end{eqnarray}

If the pair creation described by the processes (2), (3), and (4) are completed, I then have
\begin{eqnarray}
|M\rangle |\Omega\rangle\rightarrow  \frac{1}{\sqrt{2}}|M\rangle|0\rangle |0\rangle +\frac{1}{\sqrt{2}}|M-\delta m\rangle |0\rangle|1\rangle.
\end{eqnarray}

In the equation above, the blackhole is now entangled with the state of the outgoing particle: either the outside observer will \textit{not} detect a radiated particle and the blackhole has mass $M$ \textit{or} she will detect an outgoing particle and the blackhole mass is $M-\delta m$. Note that there is now an uncertainty in the mass of the blackhole since it is in a superposition of mass eigenstates $|M\rangle$ and $|M-\delta m\rangle$. Contrast this to the usual treatment of a blackhole evaporation where the blackhole is treated as a purely classical object where its \textit{exact} mass is known during the evaporation.

If the process (5) is applied $N\equiv M/\delta m$ times (think of it as a discrete-time iteration, where there is a 50\% probability of pair creation at each discrete time interval) the resulting state can be written as
\begin{eqnarray}
|\psi_N\rangle =\frac{1}{2^{N/2}}\sum_{n=0}^{N-1}\bigg[\frac{N!}{(N-n)!n!}\bigg]^{1/2}|M-n\delta m\rangle |n\rangle 
\end{eqnarray}
where $|n\rangle$ describes the state where a total of $n$ particles are radiated after the $N$th iteration. 

From the equation above, the probability that the blackhole is in mass eigenstate $|M-n\delta m\rangle$ is given by
\begin{eqnarray}
P(n)=\frac{1}{2^N}\frac{N!}{(N-n)!n!}.
\end{eqnarray}
Note that for large $N$, this probability is sharply peaked at $N/2$.

Using this expression for the probability, the average mass and the standard deviation can be calculated. Figure 1 shows the average blackhole mass versus the number of iterations or time for $N=M=1000$ and $\delta m=1$. As shown, the average mass is decreasing, that is, the blackhole is indeed evaporating. Figure 2 shows the ratio of the standard deviation and average mass $\sigma/\langle M\rangle$ versus time. Note that even in this modest scenario, $M/\delta m=10^3$, this ratio is quite small $\sigma/\langle M\rangle<0.05$ which is a consequence of the sharply-peaked probability distribution in Eq.(7). In the case of astrophysical blackhole with order-magnitude solar mass, $10^{30}$ kg, radiating a heavy particle of mass $10^{-30}$ kg(neutron/proton mass), this ratio is expected to be significantly smaller: $\sigma/\langle M\rangle\sim 10^{-60}$ for the first $10^{30}$ iterations. Hence, even though I have modelled the blackhole using a superposition of quantum states, the astrophysical blackholes still intuitively behaves like a classical one with a sharply peaked average mass. As shown in Figure 2, the said ratio is increasing and will be significantly large when $M\sim \delta m$. At this stage, the blackhole becomes a microscopic one and it is not counter-intuitive for it to be in a significant superposition of various mass eigenstates.

\begin{figure}[h!]
  \centering
     \includegraphics[width=8.25cm,height=6cm]{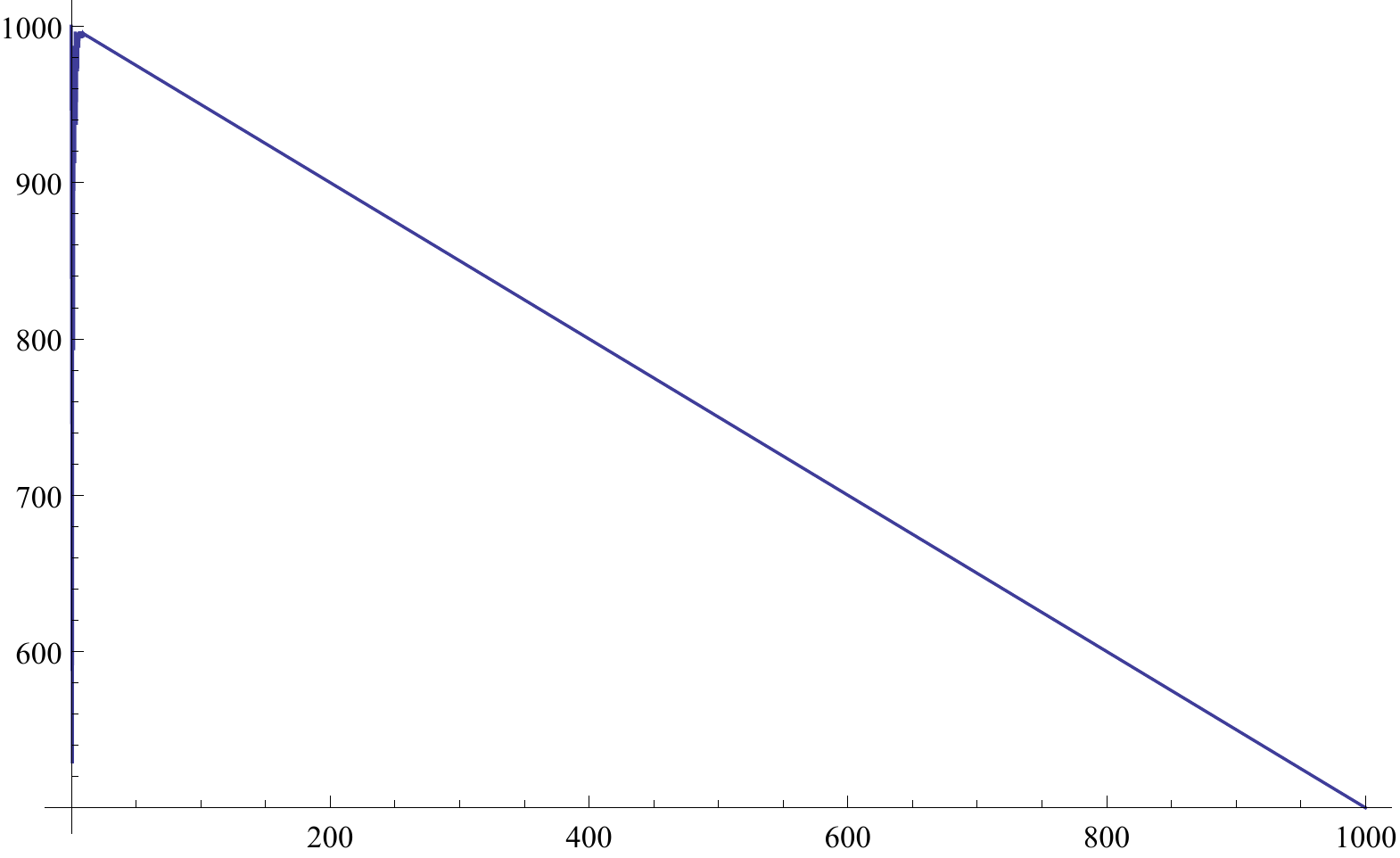}
    \caption[]{The blackhole average mass vs. time for $N=1000$ time steps. The initial mass of the blackhole is taken to be $M=1000$ while the energy of the outgoing particle is $\delta m=1$.}
\end{figure}

\begin{figure}[h!]
  \centering
     \includegraphics[width=8.25cm,height=6cm]{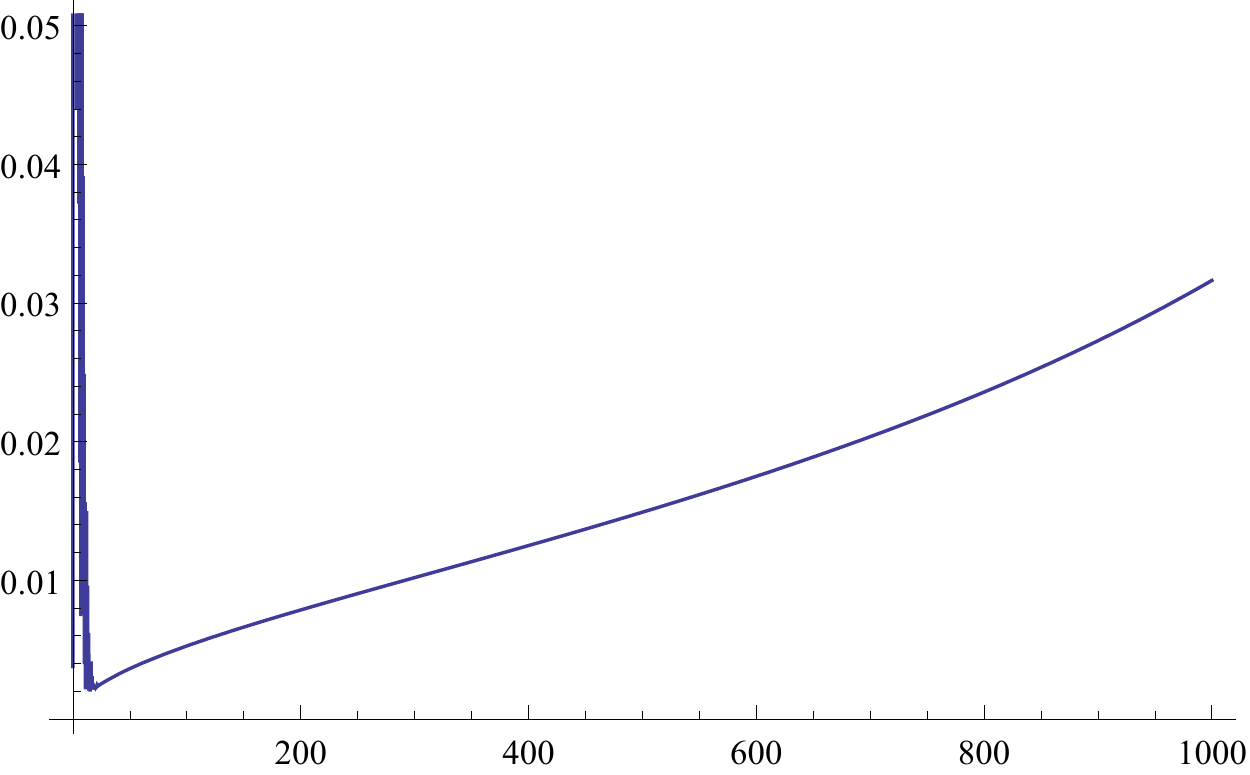}
    \caption[]{Mass tandard deviation over average mass vs. time for $N=1000$. The initial mass of the blackhole is taken to be $M=1000$ while the energy of the outgoing particle is $\delta m=1$.}
\end{figure}

\subsection{A Simple Model}
To gain further intuition about this model, I will now consider a very simple, albeit unrealistic, case where $M=\delta m=1$. After iterating $N$ times, the resulting state becomes
\begin{eqnarray}
|\psi_N\rangle =\frac{1}{2^{N/2}}|M=1\rangle |\Omega\rangle +\bigg(1-\frac{1}{2^N}\bigg)^{1/2}|M=0\rangle |1\rangle .
\end{eqnarray}

Note that the probability amplitude of the state $|M=1\rangle |\Omega\rangle$ does \textit{not} vanish in finite time. That is, unlike in classical treatment of the blackhole, one can \textit{not} say with certainty that the blackhole has fully evaporated. Although this picture might be reminiscent of the remnant blackhole scenario as a solution to the information paradox\cite{Chen15, Adler01, Aharonov87, Ellis13}, a fundamental difference must be noted. In a typical remnant scenario, the stable blackhole remnant has an \textit{exact mass}. In the scenario proposed in this paper on the other hand, the blackhole is in a superposition of various mass eigenstates so that one can not say with certainty that the blackhole has evaporated. That is, there is a residual probability that the blackhole has a non-zero mass for any finite time. The model proposed here however, does not suffer from the problem common to remnant theories where the remnant Planck-sized blackhole stores a huge amount of information.

\begin{figure}[h!]
  \centering
     \includegraphics[width=8.25cm,height=6cm]{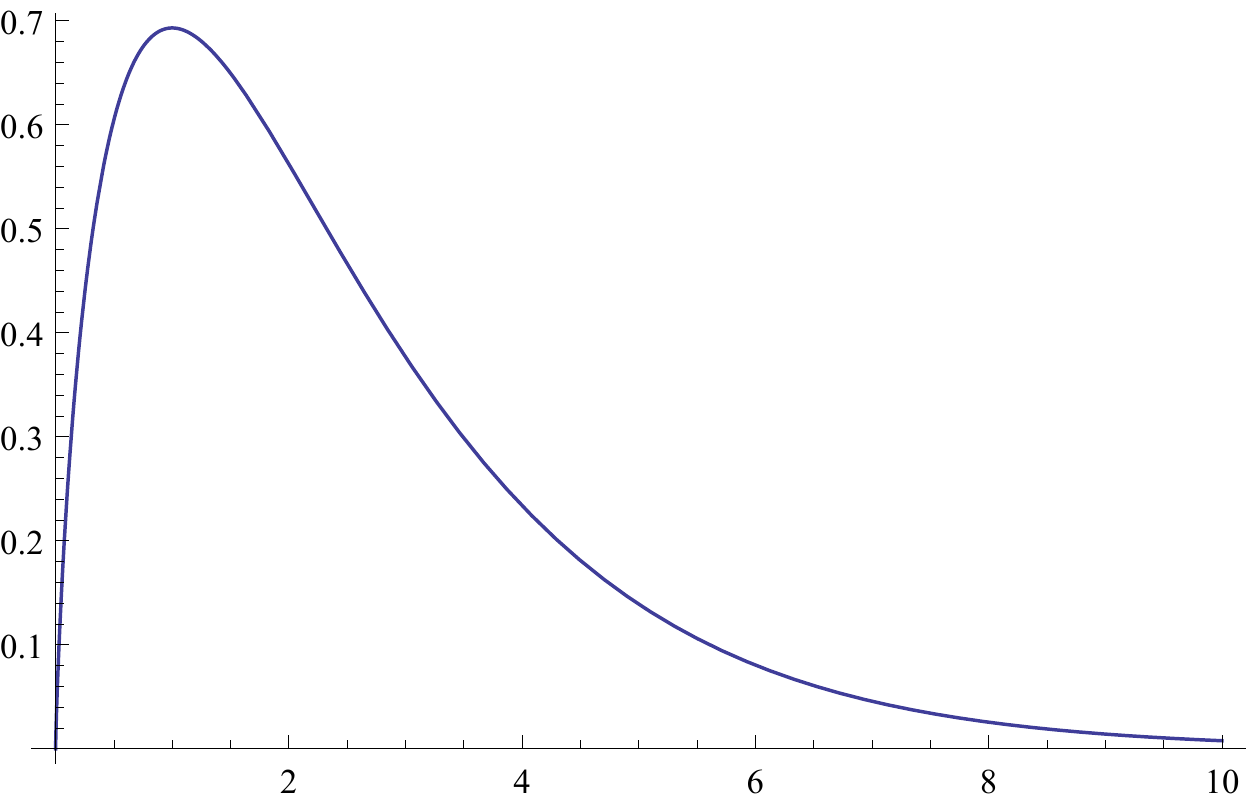}
    \caption[]{Entanglement entropy vs. time for $M=\delta m=1$.}
\end{figure}

The full density matrix corresponding to the state (8) is given by
\begin{eqnarray}
\rho(N)=
\begin{bmatrix}
    \frac{1}{2^N}      & \frac{1}{2^{N/2}}\bigg(1-\frac{1}{2^N}\bigg)^{1/2}\\
    \frac{1}{2^{N/2}}\bigg(1-\frac{1}{2^N}\bigg)^{1/2}     & 1-\frac{1}{2^N}
\end{bmatrix}.
\end{eqnarray}

It can be verified that at all times
\begin{eqnarray}
\mbox{Tr}\rho(N)^2=1
\end{eqnarray}
confirming that the evaporation process Eq.(5) used in this model is unitary. This also shows that the closed system, composed of the blackhole and the radiation, remains as a pure state at all times during the evaporation process. It is in this sense that the information paradox is solved.

The reduced density matrix can be calculated by tracing out the blackhole states in the full density matrix
\begin{eqnarray}
\rho^r=\bigg(1-\frac{1}{2^N}\bigg)|1\rangle\langle 1|+\frac{1}{2^N}|0\rangle\langle 0|.
\end{eqnarray}

This shows that the radiation is a mixed state consistent with the result of Hawking\cite{Hawking75, Hawking76}.

The entanglement entropy can then be calculated
\begin{eqnarray}
S(N)=-\bigg(1-\frac{1}{2^N}\bigg)\ln\bigg(1-\frac{1}{2^N}\bigg)-\frac{1}{2^N}\ln\frac{1}{2^N}.
\end{eqnarray}

This is plotted in Figure 3 as a function of time.

Note that after an initial increase, the entanglement entropy decreases and approaches zero as time approaches infinity. This is another demonstration that the proposed model has no information loss.

\section{Conclusion}
In this paper, it was argued that the root of the information paradox is the interaction of the classical blackhole with the quantum-mechanical radiation. A simple model where the blackhole is described by a quantum state $|M\rangle$ was proposed. It was shown that the blackhole is in a superposition of various mass eigenstates but with a sharply-peaked average mass and vanishingly small standard deviation for the case of astrophysical blackhole. This is consistent with our notion of a classical object. Using a simple realization of the model, it was shown that the closed system composed of the blackhole and its radiation, remains a pure state during the evaporation process. It was further shown, that although the entanglement entropy initially increased during the evaporation, it decreases and  asymptotically approaches zero for large times. All of this demonstrates the resolution of the information paradox.

Although the model proposed in this paper is relatively simple, the essence of the arguments can be extended to a much more realistic and far more complicated blackhole evaporation process.

\section*{Acknowledgement}
I would like to thank R. Bernardo for the useful discussions.

\end{document}